\documentstyle[aps,twocolumn]{revtex}
\begin{document}
\draft
\title{Spin-polarized transport in $GaMnAs$ multilayers}
\author{L. Loureiro da Silva, M. A. Boselli, X. F. Wang and I. C. da Cunha Lima}
\address{Instituto de F\'\i sica, Universidade do Estado do Rio de Janeiro\\
Rua S\~{a}o Francisco Xavier 524, 20.500-013 Rio de Janeiro, R.J., Brazil}
\author{and A. Ghazali}
\address{Groupe de Physique des Solides,\\
UMR 7588-CNRS, Universit\'{e}s Paris 7\\
et Paris 6\\
Tour 23, 2 Place Jussieu, F-75 251 Paris Cedex 05, France}
\maketitle
\date{\today}

\begin{abstract}
The spin-dependent mobility for the lateral transport of the hole gas in a
GaMnAs/GaAs heterostructure containing several metallic-like ferromagnetic
layers is calculated. The electronic structure is obtained self-consistently
taking into account the direct Coulomb Hartree and exchange-correlation
terms, besides the sp-d exchange interaction with the Mn magnetic moments.
\end{abstract}


Recent advances on growing $Ga_{1-x}Mn_{x}As$ multilayers opened a wide
range of interest in heterostructures of that Diluted Magnetic Semiconductor
(DMS) \cite{nat1,nat2}, specially in spin-polarized transport due to
potential applications in the area of quantum computers. Besides a magnetic
impurity, $Mn^{++}$ in this alloy is a strong $p$ dopant. However, the
density of carriers, as observed in measurements of the (anomalous) Hall
resistance, is only a small fraction (of the order of 10\%) of the $Mn^{++}$
concentration, indicating that a major part of them remains neutral. The 
{\it sp-d} exchange interaction of Ruderman-Kittel-Kasuya-Yosida (RKKY) type
has been recognized as the main origin of the observed ferromagnetism in the
metallic phase of III-V based DMS \cite{matsu,utiv}, and the possible
magnetic ordering occurring in $Ga_{1-x}Mn_{x}As$ heterostructures have been
studied {\it via} Monte Carlo simulations on metallic samples \cite{boselli}%
. Recently, a two component model has been suggested to contempt to the
iinterplay of neutral and ionized $Mn$ atoms in the magnetic interactions 
\cite{bhatt}. Chiba {\it et al} \cite{chiba} investigated a trilayer
structure and observed a ferromagnetic interaction, although weak, between
two $GaMnAs$ ferromagnetic layers. More recently, ferromagnetic arrangements
of $GaMnAs$ multilayers have been observed, forming a ferromagnetic
superlattice \cite{upsalla}. Carriers, according to their polarization, are
attracted or repelled by the magnetic layer. Therefore, the transport mean
free path is expected to be different for each polarization. The aim of this
work is to obtain the spin-polarized charge density, and to understand its
effect on the spin-polarized transport in a model structure consisting of a
sequence of $GaMnAs$ layers grown inside a thick non-magnetic $GaAs$ layer.

The spin-polarized electronic structure for holes is obtained
self-consistently in the reciprocal space \cite{ghaz}, taking into account
the hole-hole interaction as well as the hole interaction with the magnetic
impurities through the contact potential: 
\begin{equation}
U_{mag}({\bf {r})=-}I{\bf \sum_{i=1}^{N_{i}}{s}({r}).{S}({R_{i}})\delta ({r}-%
{R_{i}}),}  \label{umag}
\end{equation}
where $I$ is the $p-d$  exchange coupling constant, ${\bf {R_{i}}}$ denotes
the\ positions of the $N_{i\text{ }}$ impurities $Mn$ , ${\bf {S}({R_{i}})}$
is the (classical) spin of the impurity, and ${\bf {s}}$ is the spin of the
hole. We assume the many layer magnetizations to be oriented in a single
direction, each of them in their metallic and ferromagnetic phase. Thus, the
spin of the hole is well defined in that direction, being polarized either
up (parallel) or down(anti-parallel). When integrating the magnetic term in
the Hamiltonian over ${\bf {r}}$, the magnetic impurities are assumed to be
uniformly distributed in the $Ga_{1-x}Mn_{x}As$ DMS layers, all of them in
each magnetic layer $j$ having the same magnetization, namely the thermal
average magnetization $<{\bf {M}>_{j}}$. Therefore, 
\begin{eqnarray}
&&I\int d^{3}r\exp [i({\bf {q}-{q}^{\prime }).{r}]\sum_{i=1}^{N_{i}}{s}({r}).%
{S}({R_{i}})\delta ({r}-{R_{i}})\simeq }  \nonumber \\
&&N_{0}\beta x\frac{\sigma }{2}\sum_{j}<M>_{j}F_{DMS}^{j}(q_{z}-q_{z}^{%
\prime })(2\pi )^{3}\delta _{2}({\bf {q}_{\parallel }-{q}_{\parallel
}^{\prime }),}
\end{eqnarray}
where  $\sigma =\pm 1$ for the hole spin orientation up (parallel) or down
(anti-parallel) and $\delta _{2}$ the two-dimensional $\delta $-function in
the $q_{xy}$-plane. As usual, $N_{0}\beta =I/v_{0}$, where $v_{0}$ is the
volume of the Mn$^{++}$ ion. $F_{DMS}^{j}$ is the integral performed on the $%
z$-coordinate across the $j$-th DMS layer: 
\begin{equation}
F_{DMS}^{j}(q)\equiv \frac{1}{2\pi }\int_{DMS}^{(j)}dz\exp [iq.z].
\label{form}
\end{equation}
A net $Mn^{++}$ magnetization $<M>_{j}$ polarizes the hole gas by
introducing additional effective confining potentials given by 
\begin{equation}
V_{mag}^{eff}(z)=-N_{0}\beta x\frac{\sigma }{2}\sum_{j}<M>_{j}g_{j}(z),
\label{effecmag}
\end{equation}
where $g_{j}(z)=1$ if $z$ lies inside the $j$-layer, and $g_{j}(z)=0$
otherwise. With the use of 
\begin{equation}
U_{eff}(q)=\frac{1}{2\pi }\int dz\exp
[iq.z][U_{c}(z)+V_{mag}^{eff}(z)+U_{i}(z)],  \label{effective}
\end{equation}
where $U_{c}(z)$ and $U_{i}(z)$ are, respectively, the confining and the
hole-hole interaction potentials, one gets the one-dimensional secular
equation in the $q_{z}$-direction: 
\begin{equation}
\det \left\{ \left[ \frac{\hbar ^{2}q^{2}}{2m^{\ast }}-E\right] \delta
(q-q^{\prime })+U_{eff}(q-q^{\prime })\right\} =0.  \label{secfin}
\end{equation}
Each DMS layer works effectively as a barrier or a well for spins parallel
or anti-parallel to the local average magnetization, depending on the sign
of $N_{0}\beta $. For $x=0.05$, and for $N_{0}\beta =-1.2eV$ \cite{okaba}
this corresponds to band offsets of $\pm 30<M>_{j}meV$. Throughout the
calculation we made the hole density $p=1.\times 10^{20}cm^{-3}$, a fraction
of roughly $10$\% of the $Mn$ concentration, T=0$K$, and $<M>_{j}=5/2$. Due
to the high carrier density, several subbands are occupied.

The spin-polarized charge density distributions are shown in Fig. \ref{fig01}
for (a) one, (b) five, (c) eleven, and (d) seventeen DMS layers. We
considered symmetric structures containing a number of $Ga_{0.95}Mn_{0.05}As$
layers of width $d=20$\AA , separated $\ $by $20$\AA\ width $GaAs$ layers,
inside a $GaAs$\ QW of width  $400$\AA . A single $Ga_{0.95}Mn_{0.05}As$
layer occupying partially a $GaAs$ quantum well can be ferromagnetic for
widths smaller than the case where the DMS occupy completely the quantum
well, as it has been shown by Monte Carlo simulations \cite{jap}. However, a
single layer of $20$\AA\ is probably much too thin to be ferromagnetic. As
the number of layers increases, the spin-polarized charge density
distribution approaches that of a semiconductor superlattice with a band
offset of $\pm 75meV$. Holes with the spin polarized parallel to the average
magnetization occupy the interstitial regions between the magnetic layers,
being repelled by the $Mn$ layers due to the negative value of $N_{0}\beta $%
. Holes polarized anti-parallel to the average magnetization, on the
ccontrary, are mostly located inside the DMS layers. Their density, however,
is much lower than that of the anti-parallel spin holes, because the lowest
occupied sub-bands are for this spin orientation. We expect the
concentration of charge in the magnetic region to favor the ferromagnetic
order of the system as compared to the single-layered
structure, a fact that may be related with the recent experiments of
Sadowski {\it et al}. \cite{upsalla}

Spin motion for carriers polarized parallel and anti-parallel to the
equilibrium polarization in otherwise non-magnetic GaAs samples has been
predicted \cite{flatte} to show a difference of an order of magnitude
between the two speeds. The hole gas in these structures, when the spin-flip
scattering is neglected, can be modeled by a two-carrier system. The
corresponding Hamiltonian, with an electric field ${\bf E}=E\widehat{x}$
applied in a direction parallel to the interfaces, is separated into center
of mass (CM) and relative motions, in this case for each spin sub-system,
each one with the instantaneous position of its CM, ${\bf R}_{\sigma }(t)$,
and its drift velocity expressed in terms of the spin-dependent mobility $%
\mu _{\sigma }$, ${\bf v}_{\sigma }=\mu _{\sigma }{\bf E}$. The bare inter
(and intra) sub-band impurity scattering potential $V_{a}^{n,n^{\prime }}$
is: 
\begin{equation}
V_{a}^{n,n^{\prime }}({\bf q})=-\frac{Z_{a}e^{2}}{2\epsilon _{0}\kappa }e^{-i%
{\bf qR}_{a}}\int_{-\infty }^{\infty }dz\phi _{n}^{\ast }(z)\phi _{n^{\prime
}}(z)\frac{e^{-q|z-z_{a}|}}{q},
\end{equation}
where $\phi _{m}(z)$ is the hole wavefunction at the bottom of the sub-band $%
m$, which corresponds to a specific spin polarization. The frictional force
due to the scattering by impurities balances the net effect of the applied
electric field on the CM \cite{lei}:

\begin{equation}
f\left( v_{\sigma }\right) =\sum_{a,n,n^{\prime },{\bf q}}
|V_{a}^{n,n^{\prime }}({\bf q})|^{2} q_{x}\Pi _{2}(n,n^{\prime
},q,q_{x}v_{\sigma })= N_{\sigma }eE.
\end{equation}
Here, the screening is included through the imaginary part of the
polarization function $\Pi _{2}(n,n^{\prime },q,\omega _{\sigma })$, with
the frequency argument $\omega _{\sigma }=q_{x}v_{\sigma }$, which can be
approximated, in this simple case, by the unperturbed polarization \cite
{mahan}.

Fig. (\ref{fig02}) shows the spin polarized mobilities for a single $%
Ga_{0.95}Mn_{0.05}As$ layer in a $GaAs$ QW of width $120$\AA\ as a function
of the DMS layer width. As the width increases, corresponding to the
increase of the effective magnetic well (barrier) width, holes with spins
parallel (anti-parallel) are repelled from (attracted to) the region where
the scatterers are located. An increasing number of occupied sub-bands
corresponds to the anti-parallel polarization, reducing the parallel spin
hole charge as compared to the anti-parallel one. It follows a decrease of
the parallel polarization mobility, which is much more pronounced than that
of the anti-parallel one. The mobility for a paramagnetic sample (at T=100K)
is also included. It decreases as a function of the layer width almost in
the same way as for the majority spins. Fig. (\ref{fig03}) shows the
mobilities for a series of structures in which a variable number of $%
Ga_{0.95}Mn_{0.05}As$ layers of width $20$\AA , separated by $20$\AA\ width $%
GaAs$ layers, are symmetrically placed inside a $400$\AA\ $GaAs$ QW, as in
the calculation of the electronic structure of Fig.(\ref{fig01}). We observe
an oscillation of the mobilities for small number of layers, but the general
trend is that above a certain number of layers the mobility of the
anti-parallel polarization increase monotonically, while that of the
parallel polarization decreases, reaching a plateau, the former being a
factor of eight higher than the latter. Note that the paramagnetic mobility
reaches a plateau just above a few layers.

We obtained a strongly spin-dependent mobility, with drift velocities
differing by almost one order of magnitude, in agreement with other results 
\cite{flatte}. This has an additional interest in what concerns the hole gas
properties under an external magnetic field. On the other hand, from the
theoretical point of view of non-linear transport, one could expect that a
two-gas model presenting such a difference in the drift velocities, as shown
in this calculation, will generate higher order processes like a mutual
Coulomb drag, and other non-equilibrium effects in the two spin-polarized
electron gases.

This work was partially supported by CNPq, CAPES, FAPERJ and
CENAPAD/UNICAMP-FINEP in Brazil.

\begin{figure}[tbp]
\caption{ Spin-polarized charge density distributions for: (a) one, (b)
five, (c) eleven, and (d) seventeen $Ga_{0.95}Mn_{0.05}As$ layers of $20\AA$%
, separated by $20\AA$ width $GaAs$ layers. Solid line for total charge
density, dotted line for spin $\uparrow$ (parallel), dashed line for spin $%
\downarrow$ (anti-parallel).}
\label{fig01}
\end{figure}

\begin{figure}[tbp]
\caption{Spin-polarized and paramagnetic mobilities as functions of the
magnetic layer width for a single layer of $Ga_{0.95}Mn_{0.05}As$ inside a $%
GaAs$ non-magnetic layer of width $120\AA$.}
\label{fig02}
\end{figure}

\begin{figure}[tbp]
\caption{Spin-polarized and paramagnetic mobilities versus the number of DMS
layers for a multilayered structure of $20\AA$ width $Ga_{0.95}Mn_{0.05}As$
with a total width of $400\AA$.}
\label{fig03}
\end{figure}

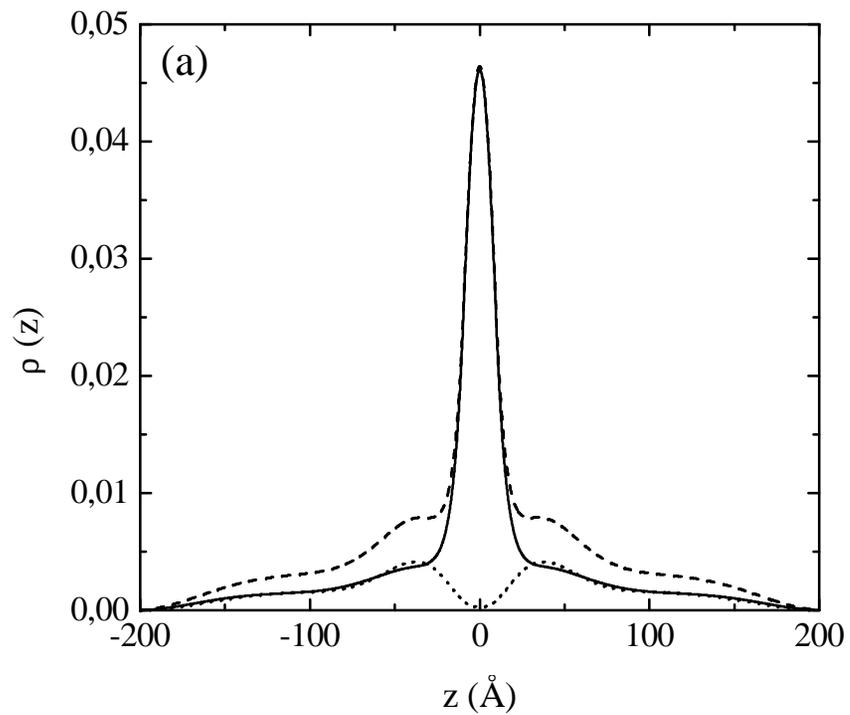
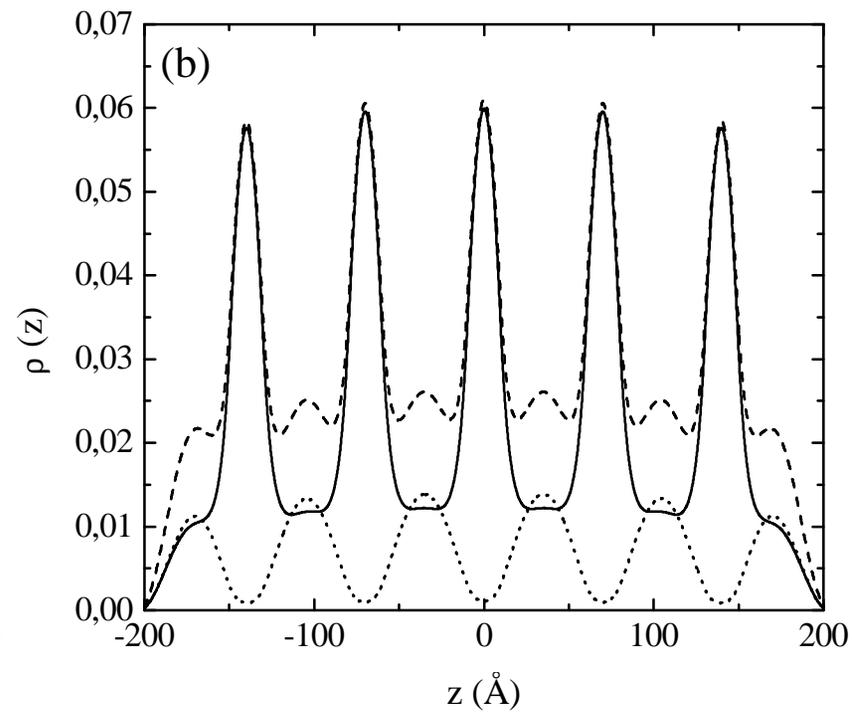
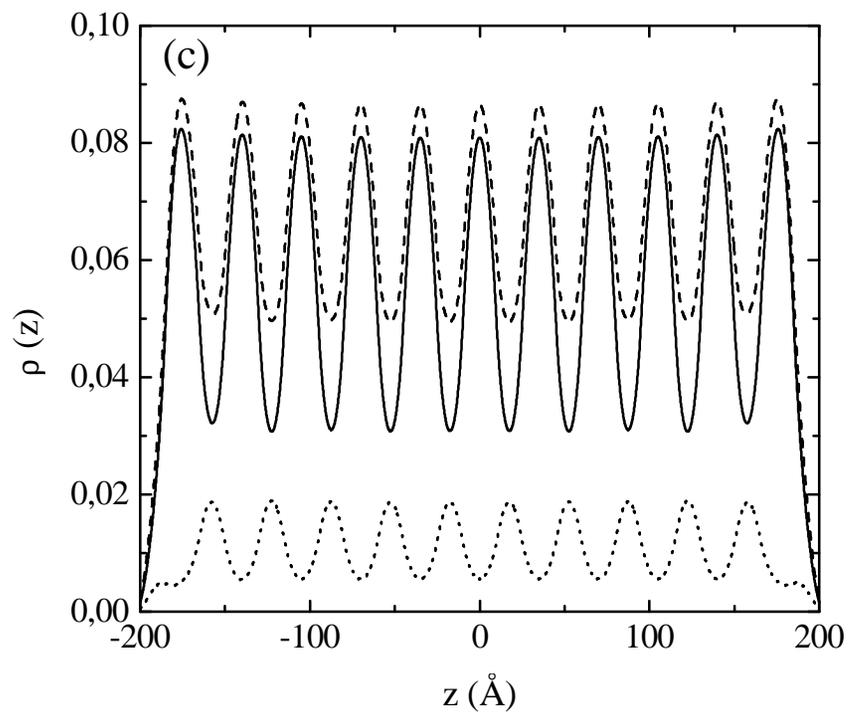
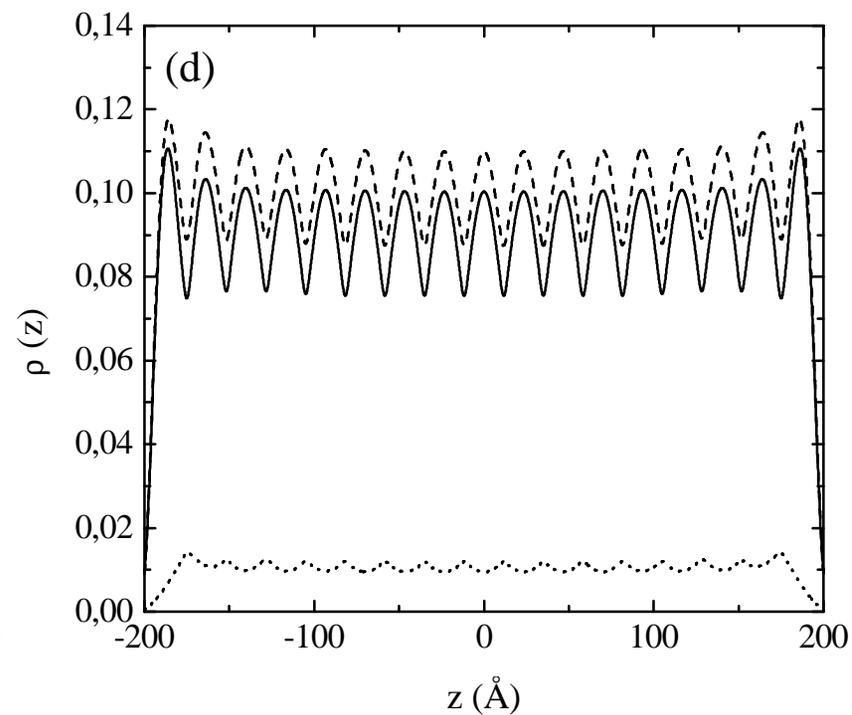
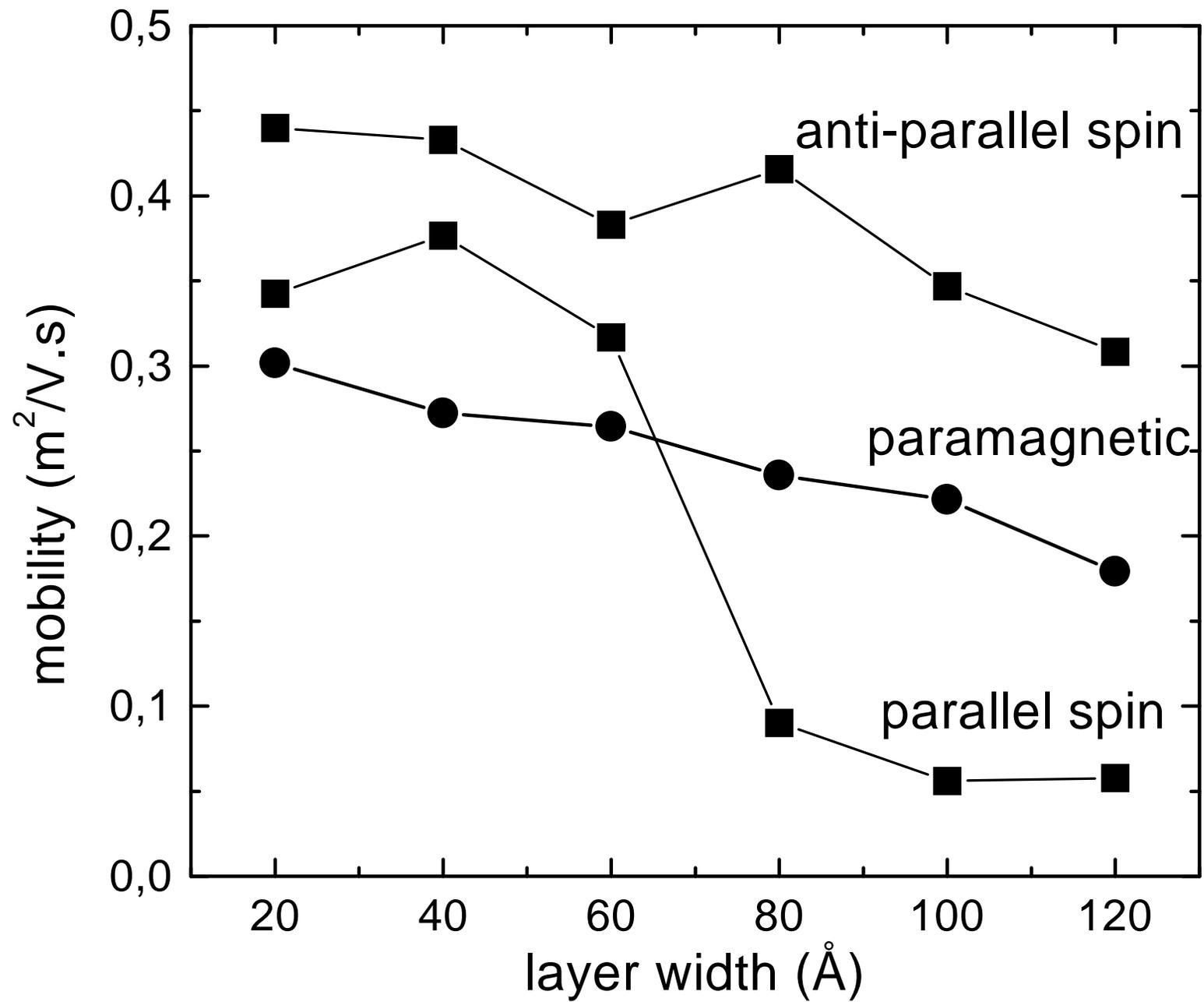
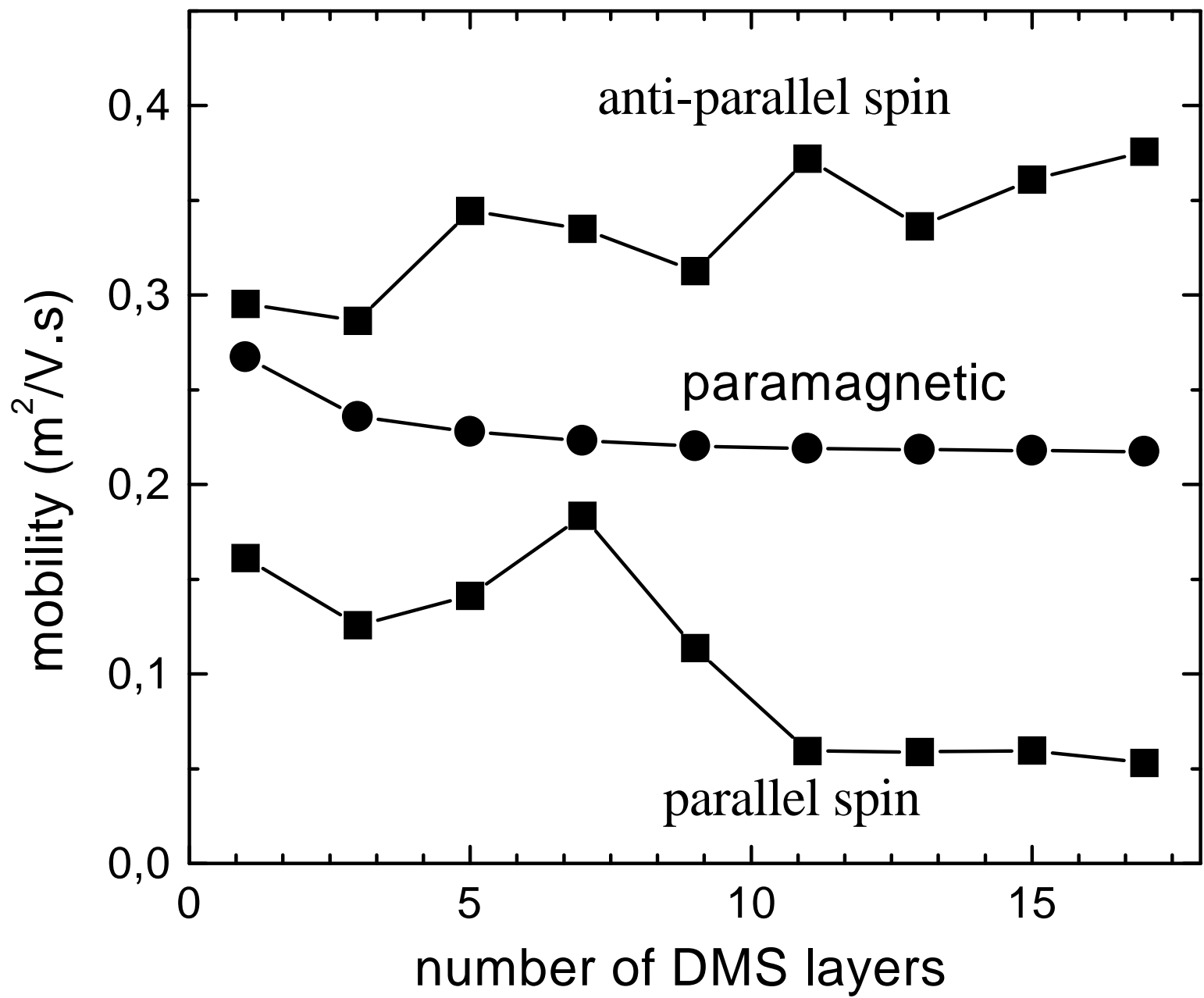
\end{document}